\documentclass[pra,aps,amsmath,amssymb,twocolumn,showpacs]{revtex4}
\usepackage[pdftex]{graphicx}
\usepackage{color}
\usepackage{amsmath}
\usepackage{amssymb}
\DeclareMathSymbol{\lang}{\mathord}{symbols}{"68}
\DeclareMathSymbol{\rang}{\mathord}{symbols}{"69}
\DeclareMathSymbol{\openbra}{\mathord}{symbols}{"68}
\DeclareMathSymbol{\closeket}{\mathord}{symbols}{"69}

\newcommand{\be}{\begin{equation}}
\newcommand{\ee}{\end{equation}}
\begin{document}
\title{The effect of Landau-Zener tunnelings on the nonlinear dynamics of 
cold atoms in  a modulated laser field} 
\author{V. Yu. Argonov}
\affiliation{Laboratory of Geophysical Hydrodynamics,
Pacific Oceanological Institute of the Russian Academy of Sciences, 
43, Baltiiskaya Street, Vladivostok, Russia 690041}

\begin{abstract}
We study the motion of a cold atom in a frequency-modulated standing laser
wave. If the detuning between the atomic electronic transition and
 the field is large,  the atom moves in a modulated optical potential 
 demonstrating known classical nonlinear effects such as
chaos and nonlinear resonances. If the atom-field
detuning is small, then two optical potentials emerge in the system, and the
 atom performs Landau-Zener (LZ) tunnelings between them. It is
a radically non-classical behavior. However, we show
that classical nonlinear structures in system's phase space (KAM-tori and 
chaotic layers) survive. Quantum effect of LZ tunnelings 
 only induces small random jumps of trajectories
between these structures (dynamical tunneling). 
\end{abstract} 

\pacs{05.45.-a, 37.10.Vz, 05.45.Mt}

\maketitle	
\section{Introduction}
Cold atoms in a laser field is a popular system for the study of quantum
 chaos. Since 1970s, a lot of authors reported various nonlinear effects in  
semiclassical models of interaction between two-level atoms and laser waves 
(constant, modulated, and kicked): dynamical chaos \cite{a48, Hensinger2000}, 
dynamical localization \cite{Graham, Moore, Graham1996}, random walking
 of atoms \cite{Konkov}, dynamical tunneling \cite{Hensinger2001,Hensinger2003},
 fractals \cite{A2003}, strange attractors \cite{A2005},
 nonlinear resonances \cite{A2006} etc. Some of these studies have not
only theoretical but also methodological aspect of quantum-classical
correspondence: they compare idealized nonlinear (in particular, chaotic)
dynamics of a semiclassical model with real dynamics of a quantum system
and analyze the difference.

As a rule, in order to observe nonlinear dynamical effects,
the model must be simple enough.  
There are two popular approaches to observe clear nonlinear dynamics and chaos
in atom-field models: (1) neglect the dynamics of atomic electronic transitions
and study only mechanical motion of an atom in a modulated laser field
(detuned far from optical resonance), (2) study the interaction between
 atomic electronical and mechanical degrees of freedom in a stationary field
 (near optical resonance). 

First approach is less fundamental, because
the system is non-autonomous. However, it requires comparatively simple 
experimental setup. Since 1990s, several important theoretical results on atomic chaos in
 a modulated field were directly tested in experiments. In particular, notable 
experiments were performed in order to observe dynamical localization \cite{Moore}
and dynamical tunneling \cite{Hensinger2001}. The phenomenon of dynamical
tunneling is very important in the framework of this paper, and we will
discuss it in details in further paragraphs. 

Second 
approach is more fundamental but methodologically more problematic
because semiclassical approximation is not very good for the description
of electronic transitions. We applied second approach in our early papers
 \cite{A2003, A2005, A2006} using idealized semiclassical models and reported a lot of nonlinear effects.
However,  
physical correctness of these results was not clear.
 In \cite{A2009,pra} more rigorous quantum analysis of atomic motion 
was performed without semiclassical approximation: atoms were considered
as wave packets, not dot-like particles.  This new model
was used to test our old results. In a "chaotic" range of
 parameters of semiclassical models \cite{A2003, A2006},
new model demonstrated random atomic Landau-Zener (LZ) transitions between two optical 
potentials. In \cite{A2009}, it was shown that LZ transitions 
(in quantum model) and dynamical chaos (in semiclassical model) 
 produce similar statistical effect on atomic mechanical motion. However, 
subtle nonlinear effects in electronic transitions seem  to
 disappear in quantum model. 

In this paper we are trying to overcome limitations of both above-mentioned
approaches and to study the nonlinear dynamics
 of an atom in a modulated standing wave near optical resonance. We use a 
stochastic trajectory model
combining semiclassical dynamics (regular atomic motion far from standing-wave
nodes) and random jumps (LZ transitions near the nodes).
We study the dynamics by the method of Poincar\'e mapping (Poincar\'e section)
and report the existence of prominent nonlinear structures (KAM-tori and chaotic areas)
 in systems's phase
 space even in presence of LZ tunnelings.  The size of some nonlinear structures is
 large enough to overcome
 Heisenberg limitations. Therefore, they seem to be not just artifacts of
 our model but a real property of the system
                         
We report not only the existence on nonlinear structures but also
LZ-induced dynamical tunneling of atom between them. 
A pair of LZ tunnelings (from one potential to another and back) may cause a jump
of atomic trajectory between two classically isolated nonlinear structures 
(e.g. two KAM-tori). Such quantum-induced jump is classically prohibited not by 
energy but by another constant of motion, and it is called dynamical tunneling. 
The phenomenon of dynamical tunneling
 was deeply studied in \cite{Hensinger2001, 
Hensinger2003} in the regime of large atom-field detunings
without LZ transitions. We report similar effect in another physical
 situation: atom is close to optical resonance, and its 
dynamical tunnelings are directly caused by pairs of successive LZ tunnelings.

\section{Equations of motion}

In this paper we study the same system as in \cite{JETPL2013, ARX2013}.
 A two-level atom with  transition frequency
 $\omega_a$ and mass $m_a$ moves in a  standing
 laser wave with modulated frequency $\omega_f[t]$.  
We neglect spontaneous emission (excited state 
has a long lifetime, or some methods are used to suppress
 decoherence).
In \cite{JETPL2013,ARX2013}, we started the analysis with 
Jaynes-Cummings Hamiltonian  obtaining quantum equations of atomic motion. 
In this paper, however, we use  a simplified stochastic trajectory model
obtained in \cite{ARX2013}. The model is consist of semiclassical and
 stochastic parts: atom moves as a dot-like particle in a classical
potential, but sometimes it performs random jumps.

\begin{figure}[htb]
\begin{center}
\includegraphics[width=0.46\textwidth,clip]{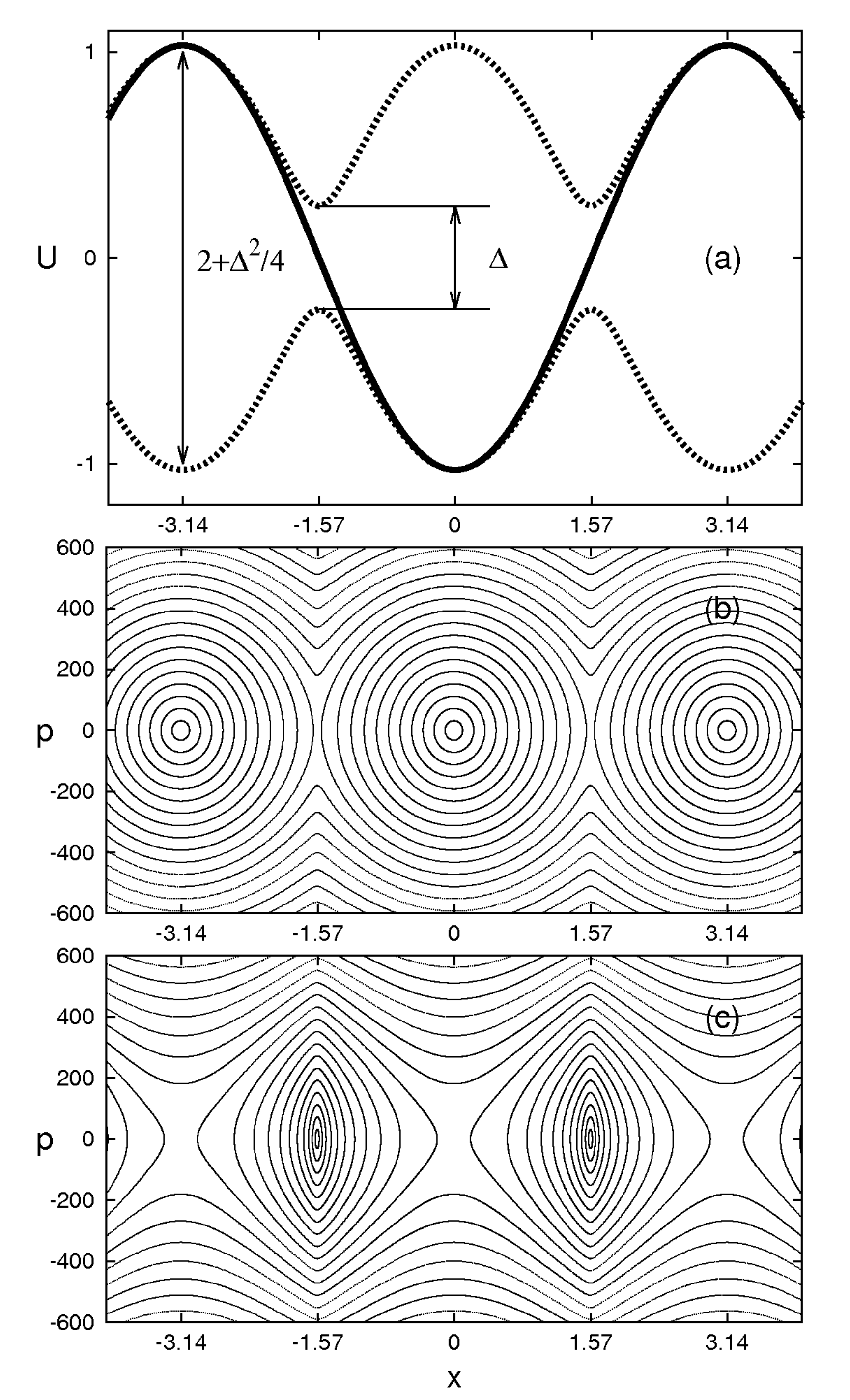}
\caption{ Optical potentials and their phase portraits: (a) dashed line: 
potentials $U^\pm$, solid line: standing wave $-\cos[x]$,
(b) phase portrait for $U^-$, (c) phase portrait for $U^+$}
\label{fig1}
\end{center}
\end{figure}

{\bf Semiclassical part}. Semiclassical approximation is valid when an atom
 moves between standing-wave 
nodes ($k_fX\ne\pi/2+\pi n $, $\cos [k_fX]\ne 0$). This motion is regular 
and may be described by the Hamiltonian
\cite{pra,A2011,JETPL2013,ARX2013}
\be
 H=\frac{P^2}{2m_a}+\hbar\Omega U^\mp, \quad U^\mp=\mp\sqrt{\cos^2[k_fX]+\frac{(\omega_a-\omega_f[t])^2}{4\Omega^2}}.
\label{U1}
\ee
Here $\Omega$ is a Rabi frequency (field intensity) and
 $k_f$ is a wave vector. 
The sign of potential $U^\mp$ is conserved, if the atom does not cross 
standing-wave nodes. 
Let us use the following normalized
 quantities: momentum  $p\equiv P/\hbar k_f$, time $\tau\equiv \Omega t$,
 position $x\equiv k_f X$,  mass  $m\equiv   m_a\Omega/\hbar k_f^2$ and
 detuning $\Delta[\tau] \equiv(\omega_f[\tau]-\omega_a)/\Omega$. 
Energy and potential (Fig.~\ref{fig1}a, dashed lines) take simple forms                                
\be
E=\frac{ p^2}{2m}+ U^\mp, \quad U^\mp=\mp\sqrt{\cos^2[x]+\frac{\Delta^2[\tau]}{4}},
\label{U}
\ee
and we obtain semiclassical equations of motion
\be
\begin{aligned}
\dot x=&\ \frac{ p}{m},\quad
\dot p=-{\rm grad} U^\mp=\frac{\sin [x] \cos [x]  }{U^\mp}.
\label{s}\end{aligned}\ee
These equations were originally obtained in \cite{A2011} for constant field
($U^\mp=U^\mp[x]$), but they also stay correct for modulated field
 ($U^\mp=U^\mp[x,\tau]$), if the modulation is slow 
 \cite{ARX2013}. Let us use 
the harmonic modulation  
\be 
\begin{aligned}
\Delta[\tau]=\Delta_0+\Delta_1\cos[\chi[\tau]],\quad \chi[\tau]\equiv\zeta \tau+\phi,\\
\zeta\ll 1,\quad \Delta_0\sim\Delta_1\ll 1. 
\label{mod} 
\end{aligned}
\ee

{\bf Stochastic part}. When an atom crosses wave nodes 
($x=\pi/2+\pi n $, $\cos[x]=0$),  the distance
between potentials is minimal (Fig.~\ref{fig1}a), and Landau-Zener (LZ)
 tunnelings between them occur with the probability \cite{A2009}
\be
W_{\rm LZ}\simeq \exp\frac{-\Delta^2 m\pi}{4 |p_{\rm node}|},
\label{LZ}
\ee
where $p_{\rm node}$ is an atomic momentum at the moment of node crossing.
For example, for $|\Delta|\ll 1$, $x[0]=0$, $U[0]=U^-$, it may be approximately
 estimated as \cite{A2011}
\begin{equation}
p_{\rm node}\simeq\pm\sqrt{p^2[0]- 2m}.
\label{pnode}
\end{equation}
If LZ tunneling occurs, the potential $U^\mp$ changes its sign and
the atom continues its motion.           
                   
As a summary, a dot-like atom moves according to
the equations (\ref{s}), but sometimes (when $\cos [x]=0$) the
 potential $U^\mp$ changes its sign with the probability (\ref{LZ}).   

\section{Phase space structure in limit cases of stationary field and 
large detunings}

In a stationary field ($\Delta_1=0$), atomic motion is simple. Atom moves in a 2D phase space: its state is
 comprehensively given by its position and momentum. In Fig.~\ref{fig1}b and c,
 phase portraits for potentials $U^\mp$ are shown. In order to compute them,
we simulated a series of trajectories with different initial positions
($x[0]=0, \pm\pi$) and momenta (a lot of values) and draw them in $x, p$
planes. All phase trajectories are
 periodic: slow atoms oscillate in potential wells (closed trajectories) and
 fast atoms  move ballistically with oscillating momentum. 
Far from  resonance ($|\Delta|\gtrsim 1$) the sign
of potential is constant, so an atom moves along the same trajectory during the evolution.
Near  resonance ($|\Delta|\ll 1$) LZ transitions between $U^-$ and $U^+$
become possible, so
the structure of phase space may switch spontaneously between Fig.~\ref{fig1}b
and Fig.~\ref{fig1}c. This may produce random walk of 
an atom (in terms of stochastic trajectory model) and splittings
of atomic wave packets (in quantum  terms) \cite{A2009, pra, A2011}.
However, this does not produce any important nonlinear effects such as dynamical
chaos. 

\begin{figure}[htb]
\begin{center}
\includegraphics[width=0.48\textwidth,clip]{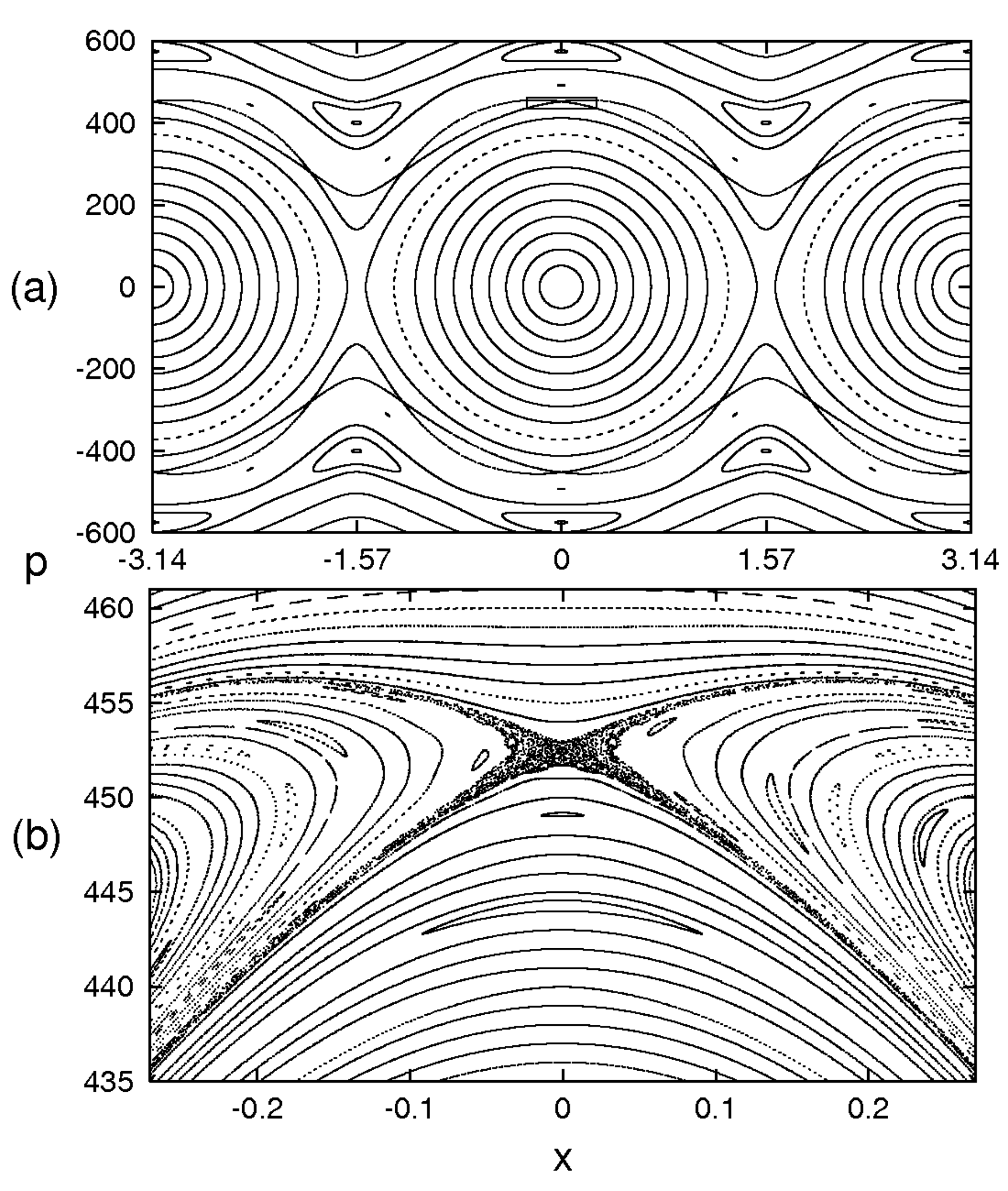}
\caption{Poincar\'e mappings in $x,p$ plane for large detuning ($\Delta_0=-0.3$):
(a) general view, (b) magnified fragment with chaotic layer and regular curves}
\label{fig2}
\end{center}
\end{figure}

Presence of field modilation (\ref{mod}) ($\Delta_1\ne 0$) radically changes
 the  dynamics. System's phase space becomes 3D because the  phase
 of modulation $\chi[\tau]$ may be regarded as a third variable. 
Far from resonance ($|\Delta|\gtrsim 1$) atom demonstrates 
 typical dynamics of 
nonlinear pendulum with periodical modulation of parameters. Similar
 pendulum regimes
were studied in countless physical systems \cite{ZS}. In order to
 demonstrate the structure phase space, let us use the method
of Poincar\'e mapping on the plane $x, p$. 
The idea of Poincar\'e mapping (Poincar\'e section) is to
draw not the whole trajectory but only its isolated points taken
periodically after each cycle of modulation. To be concrete, let us
 take points when $\chi=0$.

In Fig.~\ref{fig2}, we put $\Delta_0=-0.3$, $\Delta_1=-0.1$, $\zeta=0.02$ ($\Delta$
 oscillates in a range $-0.4\leq\Delta\leq-0.2$, so the probability of
 LZ transitions is neglible ($W_{\rm LZ}\lesssim\exp[-10]$)) and compute
Poincar\'e sections for a set of trajectories with 
different initial atomic momenta and positions in potential 
$U^-$. Most of them look similar to 2D-trajectories shown in Fig.~\ref{fig1}b,
 but some others have  sophisticated form. This picture is rather typical with
 chaotic Hamiltonian systems \cite{ZS,Z2005}. We see various regular 
Kolmogorov-Arnold-Moser (KAM) invariant
 curves (produced by nonlinear resonances of different order) and chaotic (stochastic) layers. In
 this figure, the chaotic layers are narrow
and the majority of trajectories are regular. In Fig.~\ref{fig2}b we
 magnify a small area of Fig.~\ref{fig2}a showing a saddle-like region of chaotic
 motion  between regular curves.

\begin{figure}[htb]
\begin{center}
\includegraphics[width=0.48\textwidth,clip]{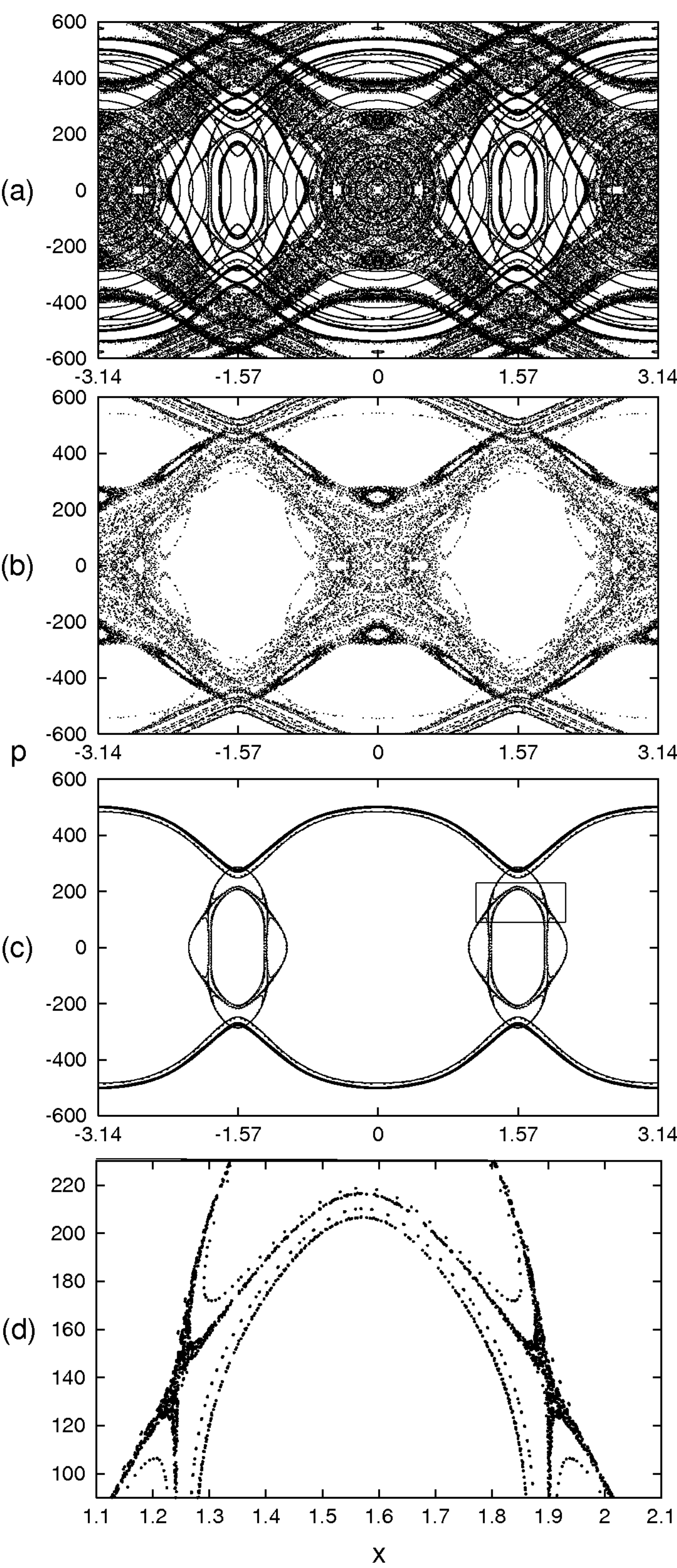}
\caption{Poincar\'e mappings in $x, p$ plane for small detuning ($\Delta_0=-0.1$):
(a) general view (many trajectories), (b) single chaotic trajectory with
$p[0]=660$, (c) single trajectory with regular and slightly chaotic
parts with $p[0]=460$, (d) magnified fragment of (c) showing chaotic layer
and neighbour regular curves visited by a single trajectory }
\label{fig3}
\end{center}
\end{figure}

Let us note that in non-autonomous system (\ref{s}), energy (\ref{U}) is
not conserved. However, for some values of initial conditions,
system contains another integral of motion (having no general 
analytical form), so the effective available phase space for particle's motion
is a sophisticated 2D surface
named KAM-tor (KAM-curves are the sections of these tori by the plane 
$\chi=0$). A particle moves periodically or quasiperiodically
on one of these surfaces. On the other hand, for some other initial conditions 
no integrals of motion exist, so the effective available phase space is 3D and
particle's motion is chaotic. The transition of particle from one KAM-tor to
 another (or to chaotic layer) is prohibited by the integral of motion. 
                                     
\section{Phase space structure in presence of modulation and Landau-Zener
transitions}

It is not surprising to see KAM-tori in a phase space of simple perturbed
nonlinear pendulum. More interesting picture is observed for smaller
detunings in  presence of LZ transitions. In Fig.~\ref{fig3}
we built Poincar\'e sections for system (\ref{s}) with the 
parameters $\Delta_0=-0.2$, $\Delta_1=-0.1$, $\zeta=0.02$. Now $\Delta$
 oscillates in a range $-0.3\leq\Delta\leq-0.1$, so $W_{\rm LZ}\lesssim \exp[-2]$, 
and LZ tunnelings  occur
many times during the evolution. The surprising fact is that 
 KAM-structures still exist. They are not destroyed by random events of
 LZ tunnelings. However, the dynamics has two new important 
 features.

First feature is that during the evolution, the potential switchs between $U^+$ and $U^-$.
Each potential corresponds to its own structure of phase space. Therefore,
 in Fig.~\ref{fig3}a
we observe two sets of overlapping structures. First set (for
$U^-$) is similar to Fig.~\ref{fig2}a (although
 chaotic regions are larger), while the second one
is a perturbed version of Fig.~\ref{fig1}c. For any single trajectory in 
Fig.~\ref{fig3}a, the motion in $U^+$ 
is comparatively slow ($|p|\lesssim p_{\rm node}$) while the motion in $U^-$
is comparativley fast ($|p|\gtrsim p_{\rm node}$). However, different
trajectories correspond to different values of $p_{\rm node}$ (\ref{pnode}). 
Therefore, it is better to study separate 
pictures for each trajectory. In Fig.~\ref{fig3}b and c we present 
Poincar\'e mappings for two particular trajectories from Fig.~\ref{fig3}a.
Here, it is easy to see structures corresponding to $U^+$ and $U^-$ without 
significant overlapping. 
                                                              
Second feature is that a single
atomic trajectory may visit many separate KAM-curves and chaotic regions. 
In absense of LZ transitions rich structures shown in Fig.~\ref{fig3}b, c
and d
can be produced only by  ensembles of trajectories with different
initial momenta. In particular, in Fig.~\ref{fig3}d we see a fine structure
of chaotic layer and regular curves. In a classical
system the transition  between them is prohibited. 
However, in presense of LZ transitions they are all produced by
 a single atomic trajectory. This is because time periods of atomic motion
between  successive node crossings in $U^{+}$ and $U^{-}$
potentials are different, and none of them (in general case) is
 synchronized with field modulation. Therefore, after two successive LZ
transitions atom returns to the same potential, but with another (in fact, random)
values of $\chi$ and $\Delta$. If
some integral of motion existed for this trajectory, it changes its value,  and the motion continues along new KAM-curve. 
In other words, a pair of  LZ tunnelings produces a dynamical tunneling
\cite{Hensinger2001,Hensinger2003} of an atomic trajectory between classically
isolated areas of phase space 

\section{Discussion and conclusion}

Two-level atom in a far-detuned modulated laser field is a popular physical
system for observation of nonlinear effects such as chaos and
nonlinear  resonance. Its phase space contains KAM-tori and chaotic layers
typical for perturbed pendulum. On the other hand, if an atom is close to optical
 resonance, it performs Landau-Zener (LZ) tunnelings between two optical
potentials, and this is not a classical pendulum behavior.
In this paper, however, we have shown that LZ tunnelings, being a quantum 
random effect, do not destroy  classical nonlinear structures in  
phase space. They affect only particular 
realizations of atomic motion producing a phenomenon of
dynamical tunneling \cite{Hensinger2001,Hensinger2003}.
Large atom-field detuning
(used in famous works \cite{Hensinger2000,Graham,Moore,Graham1996,Hensinger2001,Hensinger2003}) is not necessary
for observation of nonlinear dynamics of atoms in a modulated field.
 
These results were obtained with the use of stochastic trajectory model 
 simplifying real quantum dynamics of atoms. However, it seems
  to be correct in the framework of our study. We avoided to analyze
 phase-space structures  prohibited by Heisenberg relation. All coordinate
 scales  were or the order of optical wavelenghth (hundreds of nanometers),
 and all atomic momentums were
 much larger than the photon momentum. In \cite{ARX2013} we presented a 
 quantitative comparison between this stochastic trajectory model and purely 
quantum model, and the correspondence was good (the values of variables and parameters
 was similar to those used in this paper). 
       
The perspective of experimental test of our result
is not a simple issue. A
huge number of precise experiments with single atoms is needed. 
Each atom must be prepared in appropriate initial state and its
momentum and position must be measured after exact integer number of
 field modulation periods. In future studies we are going to simulate
such experiment numerically (using quantum equations), but this will take 
a huge computational time. 

This work has been supported by the Grant of the Russian Foundation for Basic Research
 12-02-31161.


\begin{thebibliography}{99}
\bibitem{a48} P. I. Belobrov, G. M. Zaslavskii, G. Kh. Tartakovskii, 
Zh. Eksp. Teor. Phys., {\bf 71} (1976) 1799.
\bibitem{Hensinger2000} W. K. Hensinger, A. G. Truscott, B. Upcroft, 
N. R. Heckenberg, H. J. Rubinsztein-Dunlop, J. Opt. B: Quantum Semiclass. (2000)
{\bf 2} 659.
\bibitem{Graham} R. Graham, M. Schlautmann,  and P. Zoller,  Phys. Rev. A 
{\bf 45} (1992) R19. 
\bibitem{Moore} F. L. Moore, J. C. Robinson, C. Bharucha, Bala Sundaram, and M. G. Raizen,  Phys. Rev. 
Lett. {\bf 73} (1994) 2974. 
\bibitem{Graham1996} R. Graham, S. Miyazaki. Phys. Rev. A, {\bf 53} (1996) 2683.
\bibitem{Konkov} S. V. Prants and L. E. Kon'kov, J Exp. Theor. Phys. Lett., {\bf 73} (2001) 180.
\bibitem{Hensinger2001} W. K. Hensinger, H. H\"affner, A. Browaeys,
N. R. Heckenberg, K. Helmerson, C. McKenzie, G. J. Milburn, W. D. Phillips,
S. L. Rolston, H. Rubinsztein-Dunlop, B. Upcroft, Nature, {\bf 412} (2001) 52.
\bibitem{Hensinger2003} W. K. Hensinger, N. R. Heckenberg, G. J. Milburn,
H. J. Rubinsztein-Dunlop, J. Opt. B: Quantum Semiclass. {\bf 5} (2003) R83.
\bibitem{A2003} V. Yu. Argonov and S. V. Prants, J. Exp. Theor. Phys. {\bf 96}
 (2003) 832.   
\bibitem{A2005} V. Yu. Argonov and S. V. Prants, Phys. Rev. A, {\bf 71} (2005) 053408.
\bibitem{A2006} V. Yu. Argonov and S. V. Prants, J. Russian Laser Research
{\bf 27} (2006) 360.
\bibitem{A2009} V. Yu. Argonov, J. Exp. Theor. Phys. Lett. {\bf 90} (2009) 739. 
\bibitem{pra} S. V. Prants, J. Exp. Theor. Phys. {\bf 109} (2009) 751. 
\bibitem{A2011} V. Yu. Argonov, Phys. Lett. A {\bf 375} (2011) 1116. 
\bibitem{JETPL2013} V. Yu. Argonov, J. Exp. Theor. Phys. Lett. {\bf 98} (2013) 655.
\bibitem{ARX2013} V. Yu. Argonov, Arxiv:1311.6808 (2013).                             
\bibitem{ZS} R. Z. Sagdeev, D. A. Usikov, G. M. Zaslavsky, Nonlinear Physics: 
From the Pendulum to Turbulence and Chaos. New-York: Harwood Academic
Publishers (1988).
\bibitem{Z2005} G. M. Zaslavsky, Hamiltonian Chaos and Fractional Dynamics, Oxford University Press,
Oxford (2005).

\end{thebibliography}
\end{document}